\documentclass[24pt,twoside]{article}
\usepackage[centertags]{amsmath}
\usepackage{amsfonts}
\usepackage{amssymb}
\usepackage{amsthm}
\usepackage{newlfont}
\usepackage{color}
\usepackage[svgnames,dvipsnames]{xcolor}
\usepackage{graphicx}

\date{\color{green}  2013 November}
\def \n {\noindent}
\usepackage{fancyhdr}
\usepackage[T1]{fontenc}
\usepackage[french]{babel}
\usepackage{array,multirow,makecell}
\setcellgapes{1pt}
\makegapedcells
\newcolumntype{R}[1]{>{\raggedleft\arraybackslash }b{#1}}
\newcolumntype{L}[1]{>{\raggedright\arraybackslash }b{#1}}
\newcolumntype{C}[1]{>{\centering\arraybackslash }b{#1}}

\begin{document}
\begin{center}
{\bf {\Large{\color{red}  On dynamics of wage-price spiral and stagflation in some model economic systems}}}
\end{center}
 \begin{center}
{\bf Afifa ALINTISSAR$^{1}$, Abdelkader INTISSAR$ ^{2,5}$ and Jean-Karim INTISSAR $^{3,4,5}$}
 \end{center}
\quad\\
\n $^{1}$  54, rue 260, Inara 2, Hay Moulay Abdellah, Ain-Chok \\
20470 Casablanca Morocco\\
Email address:  {\color{blue}afifa4691@gmail.com}\\

\n $^{2}$ Equipe d'Analyse Spectrale, Université de Corse, UMR-CNRS No. 6134, Quartier Grossetti\\
20 250 Corté, France\\
\n Tél: 00 33 (0) 4 95 45 00 33 -Fax: 00 33 (0) 4 95 45 00 33\\
Email address: {\color{blue} intissar@univ-corse.fr}\\

\n $^{3}$ Department of Mechanical Engineering Advance Course, Imperial College London\\
South Kensington campus London SW7 2AZ, London, United Kingdom\\
Email adress: {\color{blue}jean-karim.intissar18@imperial.ac.uk}\\

\n $^{4}$ Ecole Centrale Paris, Université  Paris-Saclay\\
Ecole Centrale Paris, 3 Rue Joliot Curie, 91190 Gif-sur-Yvette, France\\
Email adress: {\color{blue}jean-karim.intissar@student.ecp.fr }\\

\n $^{5}$ Le Prador, 129 rue du Commandant Rolland, 13008 Marseille, France\\
Email address: {\color{blue}abdelkader.intissar@orange.fr}\\
 \begin{center}
{\it Dedicated to the memory of Professor Ahmed INTISSAR (1951-2017)}
 \end{center}
  
\n  {\large{\bf {\color{red} Abstract}}}\\
 
\n  This article aims to present an elementary  analytical solution to the question of the formation of a structure of differentiation of rates of return in a classical gravitation model and in a model of the dynamics of price-wage spirals.\\

\n {\bf 2010 Mathematics Subject Classification:} B23 ; B24 ; P51\\

\n {\bf Keywords: } Classical  economic model of gravitation ; Price-wage spirals; Sraffa;  Differential system; Parameter identification method; Perron-Frobenus Theoem.\\

\n  {\large{\bf {\color{red} $\S 1$ Introduction}}}\\

\n The theory of the prices of production has the merit of having allowed a powerful criticism of a certain type of neoclassical approach, and of the theory of the corresponding value. But above all, many economists think that it can be the starting point of a positive economic theory in the mind, if not literally, of that of the classics (starting with Adam Smith and David Ricardo).\\
\n In this perspective, the prices of production relate to the long period of a capitalist economy; these are equilibrium prices for the following three reasons:\\
- they correspond to uniform rates of profit in the different sectors of the economy;\\
- they allow the equalization of the underlying offers and demands of goods;\\
- they comply with actual prices, or market prices, thus ensuring the reproduction and expansion of economic relations as a whole.\\

\n It is usually recognized that market prices differ, as a rule, from production prices. But classical economists have taken the view of Adam Smith  {\bf {\color{blue}[19]}}, who considered that the prices of production act as "centers of gravity" towards which prices.\\
However, their reasoning was not rigorous. In particular, they paid little attention to interactions between industries. \\

\n It was only in the early 1980s that this problem was approached in a more systematic way, by explicitly resorting to a dynamic process and seeking to determine when, or under what circumstances, gravitation actually takes place (cf. for example, Benetti  (1981) in {\bf {\color{blue}[2]}} and Bidard (1981) and (1986) in  {\bf {\color{blue}[3]}} and  {\bf {\color{blue}[4]}}) where we find an interpretation of the Smithian conception of market price gravitation outside the "Ricardo-Sraffaïenne" theory natural prices.\\

\n The notion of classical competition derives from the chapters of Adam Smith's "Richness of the Nations" (1976), which are devoted to the theory of prices and the theory of competition. The contemporary classical current refers to this author to elaborate a theory of alternative market prices to the representations of the General Equilibrium, (the differences with the marginalist approach are recorded in a number of works: Negishi (1985) in  {\bf {\color{blue}[15]}} and Dumenil-Levy (1982) and (1983) in  {\bf {\color{blue}[7]}} and  {\bf {\color{blue}[8]}}.\\

\n This current is the direct heir of the work of the ``Cambridgian'' current of the sixties consecrated to the counter-verses on the capital. In particular, the record of some aspects of classical thought such as the analysis of macroeconomic fluctuations and long-term trends is very disappointing and contemporaries are not up to the founders. The "unified dynamic" seen by some economists who argue that the 1980s saw a revival of classical dynamics (Smith, Ricardo and Marx) appear ambiguous must be clarified and nuanced.\\

\n For the theories of the General Equilibrium the fundamental question once demonstrated the existence of an equilibrium price vector, concerns the stability of the process which will make it possible to reach this position starting from a vector of any price (problematic of trial and error and non-trial and error).\\

\n The problem of classical competition is similar to the difference that the reference prices are so-called "natural" prices see for example Steedman in   {\bf {\color{blue}[21]}} (that for analytical convenience we equate to the production price vector of an Input-Output system). to say: if we denote by $A$ the matrix of the Input of order $n$ the vector of the natural prices is the eigenvector associated with the greatest eigenvalue of $A$ where $A$ is supposed to verify the classical hypotheses of the Perron-Frobenus theorem (see {\bf Appendix}).\\

\n Market prices fluctuate around natural prices according to the mismatches that appear between quantities supplied.\\

\n The debates of recent years have focused on the ability of the free play of competition, acting under the impetus of market forces, to resolve the imbalances over the long run, but it appears in the context of the Smithian competitive processes that the Differentiation structure of branch profit rates is assumed to be known without any economic procedure for the determination of such a structure being explicitly advanced.

\n Given that the entrepreneurs set some initial price vector $P (0) = P_{0}$ on the market, Abdelkader Intissar and Gérard Mondello sought to define differentiation rates such as convergent market prices towards production prices.\\

\n Their model is to search $(v_{1}, v _{2}, .......... , v_{n}) \in \mathbb{R}^{n}$ such that :\\
$$\displaystyle{\frac{d P}{dt} (t) = DMP(t) + f(t)}$$
\n under following conditions :\\
 $$P(0) = P_{0} \, and \,  P(T) = P^{*}$$
\n with $P(0)$ and  $f(t) = L(t) - L^{*}$  are given\\

\n where\\

\n $D$ is a diagonal matrix whose diagonal is noted $(v_{1}, v _{2}, .......... , v_{n})$,\\
\n $M$ is the matrix $B - (1+ i)A$ , given once and for all as a square matrix of order $n$ with $A$ is the matrix of inputs and $B$ is the output matrix,\\
\n $P(t)$ is the price vector of components $(P_{1}(t), P_{2} t) , ...... P_{n}(t))$, \\
\n  $\displaystyle{\frac{dP}{dt}(t)}$ denotes the derivative of the price vector,\\
 \n  $P(0)$ is the value of the price vector at point $t = 0$ ,\\
 
 \n  and  $P^{*}$  is the solution of the following system :\\
 
\n $$ \left \{ \begin{array} [c] {l}BP^{*} = (1+ i)AP^{*} + wL \\ 
\quad\\
e(B-A)P^{*} = 1 \\
\end{array} \right . $$
\n where $i$ is a given scalar and $ e $ is a unit vector of appropriate format.\\

\n In the world of Sraffa {\bf {\color{blue}[20]}}, H. Nikaido et S. Kyabashi were led to study the dynamics of Price-Wages and Stagflation Spirals. In their article {\bf {\color{blue}[16]}} (1978) they assume that the capitalists stick to a fixed rate of profit $r$ fixed and that they fix the margins on the prices according to this rate.\\

\n Price adjustment will be as follows :\\
\n $$\displaystyle{\frac{dP_{j}}{dt} (t) = (1+r) \sum_{i=1}^{n}a_{ij}P_{i}(t) + wl_{j} - P_{j},  j = 1,2, .........}$$
\n Taking into account differentiation rates, we can formulate our problem as follows:\\

\n Look for $(v_{1}, v _{2}, .......... , v_{n})$ such that :\\
\n $$\left \{ \begin{array} [c] {l}\displaystyle{\frac{dP}{dt}(t) = (1+r)A P(t) + wL - DP(t)}\\
\quad\\ 
P(0) = P_{0} \\
\quad\\
 P(T) = P^{*}\\
 \end{array} \right .  $$ \hfill { } {\bf {\color{blue} (*)}} \\
\n where $ P^{*}$ is the solution of the following system:\\
\n $$\left \{ \begin{array} [c] {l}P^{*} = (1+r)AP^{*}+ wL\\
\quad\\
eP = 1\\
\end{array}\right .$$ \hfill { } {\bf {\color{blue} (**)}}\\
\n with \\
\n $A$ is the technological matrix of inputs (supposed to be indecomposable)\\
\n $L$  is the column vector of labor requirements.\\
\n $ P^{*}$ is the column vector of production prices.\\
\n $w$ is the wage of the workers.\\
\n $r$ is the rate of profit assumed to be uniform in all branches and below the maximum rate of profit.\\
\n $D$ is a diagonal matrix whose diagonal is noted $(v_{1}, v _{2}, .......... , v_{n})$\\

\n This work aims to present an analytical solution to the question of the formation of a structure of differentiation of rates of return in a classical model of gravitation and in a model of the dynamics of price-wage spirals.\\

\n {\large{\color{red}$\S$ {\bf  2 An analytical study}}}\\

\n The two differential systems presented in the introduction are not classic in that we will not try to determine only a solution starting from a given initial position $P(0)$.\\

\n The question that concerns us is the following :\\

\n Knowing that the entrepreneurs fix on the market a vector of any initial price $P(0) = P_{0}$, one seeks to define the differentiation rates such as the convergent market prices towards the production prices.\\

\n {\bf {\color{blue}Remark 2.1}}\\

\n a) If we give ourselves the differentiation rates, we can easily solve our two differential systems.\\

\n b) The question asked in this work is whether we can determine $ V = (v_{1}, v _{2}, .......... , v_{n})$ such that  the market prices converge to the production price for $T$ given. \\

\n Mathematically, the problem is to look for $ V$ such that $P(T,V) = P^{*}$  where $P(t,V)$ is a solution of one of the two systems. \\

\n Immediately, we give a counter-example on the question of the existence of a solution :\\

\n Consider the following problems in $\mathbb{R}$:\\

\n $$\frac{dP}{dt}(t) = vaP(t)$$\\
 with $P(0)$ is given and \\
\n $$\frac{dP}{dt}(t) =(a-v)P(t)$$ \\
with $a > 0$ \\

\n For fixed $P^{*}$, the solution  of first system satisfies the following constraint :\\
\n $$P(0)e^{vaT} = P^{*}$$\\
 \n By choosing $P(0)$ and  $P^{*}$ of contrary sign, it is clear that the problem does not admit of solution !\\

\n Moreover, the positivity of $P(0)$ and $P^{*}$ is not sufficient to guarantee the positivity of $v$ and in this case we have : \\
\n $$\displaystyle{v = \frac{1}{Ta} ln(\frac{P^{*}}{P(0)})}$$\\
\n  So just choose $P(0) > P^{*}$ to infer the non-existence of an economically viable solution.\\

\n When $P^{*}$ is given, the solution of the second system satisfies the following constraint :\\
\n $$P(0)e(a-v)T = P^{*}$$\\
\n and by choosing  $P(0)$ and  $P^{*}$ of contrary sign, it follows  that the problem does not admit of solution !\\

\n {\bf {\color{blue}Remark 2.2}}\\

\n a) By taking $ v $ as unknown is essential for differentiation rate theory to make sense.\\

\n b) The resolution of our systems consists to combine two classical methods, the Runge-Kutta method for the resolution of a linear differential system and the Newton-Kantorovitch method.\\

\n Runge-Kutta formulas are the most used for solving a differential system because :\\

\n - They are easy to program, \\
- They are generally stable, \\
- The width of the step can be changed without difficulty, \\
- Above all, they "start" by themselves: the knowledge of $ P (0) $ suffices to integrate the differential system.\\
\n  We can refer for a systematic study of this method to  Kaps-Rentrop in  {\bf {\color{blue}[13]}} and for a complete study of Newton's method we refer for example to the work of the academician L. Kantorovich, and in particular to his book with G. Akilov (1981) {\bf {\color{blue}[12]}}.\\

\n {\large{\bf {\color{red} $\S$ 3 Study of the convergence of market prices to production prices}}}\\

\n In this section we discuss the resolution of following equation :\\
\n $$P(T,V)  = P^{*}$$ \\
\n Let $G(V) = P(T,V) - P^{*}$ with $G(V) \in \mathbb{R}^{n}$ and trying to find $u$  such that $G(u) = 0$.\\

\n To solve this system, we apply the Newton-Kantorovitch method. We will calculate the Fréchet derivative of $ G $ which we will note: \\
\n $$\displaystyle{H(V) = [\frac{\partial G_{i}}{\partial v_{j}}(V)] = (H_{ij})_{1 \leq i, j \leq n}}$$\\
\n Since $\displaystyle{G(V) = P(T, V) - P^{*}}$, we deduce that :\\
\n $$\displaystyle{(H_{ij} = \frac{\partial P_{i}}{\partial v_{j}})_{1 \leq i, j \leq n}}$$\\
\n We write our {\bf first system} as follows :\\
\n $$\displaystyle{\frac{dP_{i}}{dt}(t, v_{1}, v_{2}, ....., v_{n}) =  v_{i}\sum_{k=1}^{n}a_{ik}P_{k}(t, v_{1}, v_{2}, ....., v_{n}) + f_{i}(t)}$$\\
\n with $i \in [1, n]$.\\

\n for fixed $i$, if we derive the above equation with respect $v_{j}$; $j \in [1, n]$, we get a {\bf second system}:\\

\n for $j \neq i$ :\\
\n $$\displaystyle{\frac{d}{dt}[\frac{\partial P_{i}}{\partial v_{j}}](t, v_{1}, v_{2}, ....., v_{n}) =  v_{i}\sum_{k=1}^{n}a_{ik}\frac{\partial P_{k}}{\partial v_{j}}(t, v_{1}, v_{2}, ....., v_{n})}$$\
\n and for $j = i$: \\
\n $$\displaystyle{\frac{d}{dt}[\frac{\partial P_{i}}{\partial v_{i}}](t, v_{1}, v_{2}, ....., v_{n}) = \sum_{k=1}^{n}a_{ik}\frac{\partial P_{k}}{\partial v_{j}}(t, v_{1}, v_{2}, ....., v_{n}) +  v_{i}\sum_{k=1}^{n}a_{ik}\frac{\partial P_{k}}{\partial v_{i}}(t, v_{1}, v_{2}, ....., v_{n})}$$
\n By grouping the starting system and the one we have just written, we obtain a differential system of $n (n + 1)$  equations in following form :\\
\n  $$\displaystyle{\frac{d\mathbb{P}_{1}}{dt}(t, v) = \mathbb{M}_{1} \mathbb{P}_{1}(t, v) +\mathfrak{f}_{1}(t)}$$
\n with $\mathbb{P}_{1}(0)$ is fixed.\\

\n $\mathbb{M}_{1}$ is the matrix of order $n(n+1)$ defined  by grouping  the two systems,\\
\n $$\displaystyle{\mathbb{P}_{1}(t) =   (P_{1}(t), P_{2}(t) \quad . \quad. \quad. \quad P_{n}(t),  \frac{\partial P_{1}}{\partial v_{1}}(t), \frac{\partial P_{2}}{\partial v_{2}}(t), . \quad . \quad . \quad. \quad\frac{\partial P_{n}}{\partial v_{n}}(t))}$$
 and \\
 $$\displaystyle{\mathfrak{f}_{1}(t) =  ( f_{1}(t),  f_{2}(t) .  . . .  f_{n}(t), 0,  0. . . . .  0)}$$\\
\n We solve this system by the Runge-Kutta method of order $ 4 $ which give us the solution at $t = T$ and we deduce $G(v)$ et $H(v)$.\\

\n At the end of the resolution of the differential system we will have :\\
\n $$\displaystyle{\mathbb{P}_{1}(T) =  ( P_{1}(T),  P_{2}(T), .... ,P_{n}(T), \frac{\partial P_{1}}{\partial v_{1}}(T),  \frac{\partial P_{2}}{\partial v_{2}}(T), ....., \frac{\partial P_{n}}{\partial v_{n}}(T))}$$
\n Now, let $v_{0}$ be a vector in $\mathbb{R}^{n}$,  by solving the last system we obtain $P(T,v_{0})$ and $H(v_{0})$  and we define $v_{1}$ as solution  of the following system  :\\
\n $$H(v_{0})(v_{1} - v_{0}) = - G(v_{0})$$
\n where $G(v_{0}) = P(T, v_{0}) - P^{*}$\\
\n Iteratively we have :\\
\n  $$H(v_{k})d_{k}  = - G(v_{k})$$
\n  where $d_{k} = v_{k+1} - v_{k}$ is the descent direction used in Newton's classical method.\\

\n As $H (v)$ (in general case) has no particular property, we use the Gauss method with search of the pivot to solve the last system..\\

\n {\bf {\color{blue}Remark 3.1}}\\

\n a) The solution of our {\bf first system} is given by: \\
\n $$\displaystyle{P(t) = e^{tDM}P(0) + \int_{0}^{t}e^{(t - s)DM}f(s)ds}$$
\n It follows that:\\
\n $$\displaystyle{P^{*} = P(T) = e^{TDM}P(0) + \int_{0}^{T}e^{(t -s)DM}f(s)ds}$$

\n Suppose that $B = I$ , $L(t) = L  \, \, \forall \, t$ and that the real parts of the eigenvalues of $I- (1+i)A $ are negative, then $P(t)$ converges to $P^{*}$ as $t \longrightarrow + \infty$.\\

\n As $M = B - (1+i)A$, the eigenvalues of $M$ can not be characterized without imposing restrictive conditions on the matrices $A$ and $B$.\\

\n b) The method, which consists of combining the Runge-Kutta method  {\bf {\color{blue}[13]}} and that of Newton  {\bf {\color{blue}[12]}}, allows us to study similarly the second system.\\

\n Let a diagonal matrix $\displaystyle{D = (d_{ij}) ; d_{ij} = v_{i} \delta_{ij}}$ where $\delta_{ij} = 0 $ if $i \neq j$ and $\delta_{ii} = 1$ with $ i,j = 1, 2, ........ , n$\\

\n Find $(v_{1} ,v_{2} , ...... , v_{n})$ such that\\
\n $$\left \{ \begin{array} [c] {l}\displaystyle{\frac{dP}{dt}(t) = (1+r) AP(t) - DP(t) + f(t)}\\ 
\quad\\
P(0) = P_{0}\\
\quad\\
P(T) = P^{*}\\
\end{array} \right . $$
\n Our parameter identification method can be used for other families of differential systems.\\

\n {\large {\bf {\color{red}Appendix :}}}\\

\n In 1907, Perron  {\bf {\color{blue}[17]}} and  {\bf {\color{blue}[18]}} gave proofs of the following famous theorem, which now bears his name, on positive matrices:\\
 
\n {\color{blue}{\bf Theorem A}} (Perron Theorem)\\

\n Let $A$ be a square positive matrix. Then $\rho(A)$ is a simple eigenvalue of $A$ and there is a corresponding positive eigenvector.\\
Furthermore,    $\mid \lambda \mid < \rho(A)$ for all $\lambda \in \sigma(A), \lambda \neq \rho(A)$.\\

\n Here we denote by $\sigma(A)$ the spectrum (the set of all eigenvalues) of a (square) matrix $A$, and by $\rho(A)$ the spectral radius of $A$, i.e., the quantity $max \{ \mid \lambda \mid ;  \lambda \in \sigma(A)\}$. By a positive (respectively, nonnegative) matrix we mean a real matrix each of whose entries is a positive (respectively, nonnegative) number.\\
\n  If $A$ is a nonnegative (respectively, positive) matrix, we shall denote it by $A \geq 0$ (respectively, $A > 0$).\\

\n {\color{blue} {\bf Definition B}} (reducible matrix respectively irreducible matrix)\\

\n An $n \times n $ matrix $A = (a_{ij}) $ is said to be reducible if $n \geq 2$ and there exists a permutation matrix P such that :\\
\n $$^{t}PAP = \left [ \begin{array} {cc} A_{11}&A_{12}\\
\quad\\
0&A_{22}\\
\end{array} \right ]$$
\n where  $A_{11}$ and $A_{12}$ are square matrices of order at least one. If A is not reducible, then it is said to be irreducible.\\

\n In $1912$, Frobenius  {\bf {\color{blue}[9]}} extended the Perron theorem to the class of irreducible nonnegative matrices:\\

\n {\color{blue}{\bf Theorem C}} (Frobenius Theorem)\\

\n Let $A \geq 0$ be irreducible. Then \\

 (i)  $\rho(A)$ is simple eigenvalue of $A$, and there is a corresponding positive eigenvector.\\

 (ii) If $A$ has $m$ eigenvalues of modulus $\rho(A)$, then they are in the following form $\displaystyle{\rho(A)e^{\frac{2ik\pi}{m}}}$ ; \\
 .\quad \quad \quad  $ k = 0, ...., m-1$.\\

  (iii) The spectrum of $A$ is invariant under a rotation about the origin of the complex plane\\
   .\quad \quad \,\quad  by $\frac{2\pi}{m}$, i.e., $\displaystyle{e^{\frac{2i\pi}{m}}\sigma(A) = \sigma(A)}$.\\

 (iv) If $m > 1$ then there exists a permutation matrix $P$ such that :\\
\n $$^{t}PAP = \left (\begin{array}{ccccc} 0&A_{12}&&&\\
\quad\\
&0&A_{23}&&\\
\quad\\
 & &0&\ddots&\\
\quad\\
&&&\ddots&A_{m-1,m}\\
\quad\\
A_{m,1}&&&&0\\
\end{array} \right )$$
\quad\\
\n where the zero blocks along the diagonal are square.\\

\n we refer the reader to the intersting papers of C. Bidard and M. Zerner  {\bf {\color{blue}[5]}},  {\bf {\color{blue}[6]}} for an application of the Perron-Frobenus theorem in relative spectral theory.\\

\n A natural extension of the concept of a nonnegative matrix is that of an integral operator with a nonnegative kernel. The following extension of Perron?s theorem is due to Jentzsch {\bf{\color{blue}[11]}}:\\

\n {\color{blue}{\bf Theorem D}} (Jentzsch Theorem)\\

\n Let $k(., .)$ be a continuous real function on the unit square with $k(s, t) > 0$ for all $0 \leq s, t \leq 1$. If $K: L^{2}[0, 1] \longrightarrow  L^{2}[0, 1]$ denotes the integral operator with kernel $k$ defined by setting \\
$$ (Kf)(s) = \int_{0}^{1}k(s, t)f(t)dt , f \in L^{2}[0, 1],$$

\n then\\

(i) $K$ has positive spectral radius;\\

(ii) the spectral radius $\rho(K)$ is a simple eigenvalue, with (strictly) positive eigenvector;\\

(iii) if $ \lambda = \rho(K)$ is any other eigenvalue of $K$, then $\mid \lambda \mid < \rho(K)$.\\

\n {\color{blue}{\bf Remark E}}\\

\n (ii) We refer the reader to an original application of Jentzsch Theorem in reggeons field theory  see T. Ando and M. Zerner $(1984)$ {\bf {\color{blue}[1]}} and A. Intissar $(1989)$ {\bf {\color{blue}[10]}} \\

\n (ii) In $1948$, in an abstract order-theoretic setting, in the important memoir {\bf{\color{blue}[14]}} Krein and Rutman extended the theory to a compact linear operator leaving invariant
a convex cone in a Banach space.\\

\n They obtained the following:\\
 
\n {\color{blue}{\bf Theorem F}} (Krein-Rutman Theorem)\\

\n Let $A$ be a compact linear operator on a Banach space $\mathbb{X}$. Suppose that $A(\mathcal{C}) \subseteq \mathcal{C}$, where $\mathcal{C}$ is a closed generating cone in
$\mathbb{X}$. If $\rho(A) > 0$, then there exists a nonzero vector $x \in \mathcal{C}$ such that $Ax = \rho(A)x$.\\

\n {\large {\bf {\color{red}References}}}\\

\n {\bf {\color{blue}[1]} }T. Ando and M. Zerner :. Sur une Valeur propre d'un opérateur, Commun. Math. Phys. 93, (1984) 123-139\\

\n {\color{blue}{\bf [2]}} C. Benetti :. La question de la gravitation des prix de marché dans La richesse des nations, Cahiers d'Économie Politique Année, volume 6, (1981) pp. 9-31\\

\n {\color{blue} {\bf [3]}} C. Bidard :.  Travail et salaire chez Sraffa , Revue Economique, vol.33,  (1881) pp. 365-373\\

\n {\color{blue}{\bf [4]}} C. Bidard  :. Baisse tendancielle du taux de profit et marchandise étalon, Economie Appliquée, vol.39, (1986) pp.139-154.\\

\n {\color{blue}{\bf [5]}} C. Bidard  et  M. Zerner :. Positivité en théorie spectrale relative », Comptes Rendus de l'Académie des Sciences, vol.310,  série I, (1990) pp.709-712.\\

\n {\color{blue} {\bf [6]}} C. Bidard and Zerner M. :.  The Perron-Frobenius theorem in relative spectral theory », Mathematische Annalen, vol.289,  (1991) pp.451-464. \\

 \n {\color{blue}{\bf [7]}} G. Duménil and D. Lévy :.  "Valeur et prix de production, le cas des productions jointes", Revue économique, Vol. 33 (1),  (1982) pp. 30-70.\\
 
 \n {\color{blue}{\bf [8]}} G. Duménil and Lévy :. "Prix et Quantités: le cas des productions jointes", Économie appliquée, Vol. 34  (2), (1983) pp. 411-445 \\
 
 \n {\color{blue}{\bf [9]}} G. F. Frobenius :. Uber Matrizen aus nicht negativen Elementen, Sitzungsber. Kon. Preuss. Akad. Wiss. Berlin, (1912), 456-477\\
 
 \n {\color{blue} {\bf [10]}} A.  Intissar :.  Quelques nouvelles propriétés spectrales de l'hamiltonien de la théorie des champs de reggeons. C. R. Acad. Sci. Paris, Série I, 308, (1989)\\
 
\n {\color{blue} {\bf [11]}} R. Jentzsch :. Uber Integralgleichungen mit positiven Kern, J. Reine Angew.Math. 141 (1912), 235-244.\\

\n {\color{blue}{\bf [12]}} L. Kantorovitch et  G. Akilov  :.  Analyse fonctionnelle Tome I et Tome II , Editions MIR (1981)\\

\n  {\color{blue}{\bf [13]}} P.  Kaps and P. Rentrop :. Generalized  Runge-Kutta  Methods  of  Order  Four  with  Stepsize  Control  for  Stiff  Ordinary Differential Equations, Numerische Mathematik, 33, 55-68, (1979)\\

\n {\color{blue} {\bf [14]}} M. G. Krein and M. A. Rutman :.  Linear operators leaving invariant a cone in a Banach space, Amer. Math. Soc. Transl. Ser. 1 10 (1950), 199-325 [originally Uspekhi Mat. Nauk 3 (1948), 3-95].\\
 
 \n {\color{blue}{\bf [15]}} T. Negishi :.  Economic  theories  in  a  nonwalrasian  tradition Cambridge University Press, New York, (1985)\\
 
\n {\color{blue} {\bf [16]}} H. Nikaido and S. Kyabashi :. Dynamics of wage-price spiral and stagflation in the Leontieff- Sraffa system, International Economic Review, (1978).\\

\n  {\color{blue}{\bf [17]}} O. Perron :. Grundlagen fur eine Theorie des Jacobischen Kettenbruchalgorithmus, Math. Ann. 63 (1907), 1-76.\\

\n {\color{blue}{\bf [18]}} O. Perron :.  Zur Theorie der uber Matrizen, Math. Ann. 64 (1907), 248-263.\\

\n {\color{blue} {\bf [19]}} A. Smith :.  An  Inquiry into  the  Nature    and  Causes  of  the  Wealth  of  Nations, edited by E. Cannan (Chicago: The University of Chicago Press)  (1976). \\
 
  \n {\color{blue}{\bf [20]}} P. Sraffa :. Production of commodities by means of commodities, Cambridge University Press (1960).\\
    
\n {\color{blue}{\bf [21]}} I. Steedman :. Natural prices, differential profit rates and the classical competitive process. The Manchester School of Economic and Social Studies 52, ( 1984) pp.123-140.\\

 \end{document}